# Teaching LLMs Music Theory with In-Context Learning and Chain-of-Thought Prompting: Pedagogical Strategies for Machines

Liam Pond[a] and Ichiro Fujinaga[b]
*Schulich School of Music, McGill University, Montréal, Canada*
liam.pond@mail.mcgill.ca, ichiro.fujinaga@mcgill.ca

Abstract: This study evaluates the baseline capabilities of Large Language Models (LLMs) like ChatGPT, Claude, and Gemini to learn concepts in music theory through in-context learning and chain-of-thought prompting. Using carefully designed prompts (in-context learning) and step-by-step worked examples (chain-of-thought prompting), we explore how LLMs can be taught increasingly complex material and how pedagogical strategies for human learners translate to educating machines. Performance is evaluated using questions from an official Canadian Royal Conservatory of Music (RCM) Level 6 examination, which covers a comprehensive range of topics, including interval and chord identification, key detection, cadence classification, and metrical analysis. Additionally, we evaluate the suitability of various music encoding formats for these tasks (ABC, Humdrum, MEI, MusicXML). All experiments were run both with and without contextual prompts. Results indicate that without context, ChatGPT with MEI performs the best at 52%, while with context, Claude with MEI performs the best at 75%. Future work will further refine prompts and expand to cover more advanced music theory concepts. This research contributes to the broader understanding of teaching LLMs and has applications for educators, students, and developers of AI music tools alike.

## 1 INTRODUCTION

The recent emergence of Large Language Models (LLMs) like ChatGPT, Claude, and Gemini, is revolutionizing traditional norms, workflows, and even social structures (Evron & Tartakovsky, 2024). From automated healthcare to improved agricultural yields, nearly every aspect of our world is changing, and education is no exception (Raiaan et al., 2024).

As flipped classroom and active learning strategies continue to gain traction in pedagogy, teachers are increasingly turning to the assistance of digital platforms. Given that effective use of these techniques has been shown to increase student independence, motivation, curiosity, and confidence (Hui et al., 2021), LLMs present a promising opportunity to take this approach to the next level.

Already, LLMs are transforming education by generating lesson plans, quizzes, and practice problems that are tailored to the needs of individual students (Hu et al., 2024; Kazemitabaar et al., 2024). Additionally, they allow teachers to dedicate more time to meaningful engagement with students by serving as teaching assistants that streamline the grading and feedback process (Alsafari et al., 2024; Liu et al., 2025). Moreover, LLMs are an inclusive and accessible resource that learners can access at any time. In this way, they can help mitigate systemic inequalities by providing free multilingual support to students who may not have the means for additional resources like private tutors, levelling the playing field for those with diverse socioeconomic backgrounds (İlhan et al., 2024).

However, despite being effective at teaching, evaluating, and explaining many diverse concepts, LLMs currently lack much of the specialized knowledge required for pedagogical music theory applications. Furthermore, existing research on music-related applications of LLMs has focused on tasks in music generation (Deng et al., 2024; Kharlashkin, 2024), recommender systems (Sakib & Bijoy Das, 2024), and even music curriculum design

[a] https://orcid.org/0009-0003-0554-1893
[b] https://orcid.org/0000-0003-2524-8582

(Shang, 2024), leaving music theory and analysis relatively unexplored.

To address the gap in the application of LLMs to music theory, this study conducts exploratory research into teaching LLMs music theory using strategic prompting, focusing on two specific techniques: in-context learning and chain-of-thought prompting (CoT).

In-context learning refers to the process of leveraging large prompt context windows by providing instructions, guides, and examples of solutions to problems LLMs would otherwise not be able to understand. This approach allows LLMs to identify and apply new patterns without requiring additional fine-tuning or modifications to the underlying model (Dong et al., 2024).

In a similar vein, chain-of-thought prompting improves performance on reasoning and logic-based problems by guiding them to articulate intermediate steps, breaking down complex problems into smaller, simpler tasks. By providing examples of the correct chain of thought, LLMs can be taught to logically process unfamiliar concepts, mimicking human problem-solving.

As emerging strategies in prompt design, in-context learning and chain-of-thought prompting hold significant potential for enhancing music theory comprehension in LLMs, yet their application in this domain remains largely unexplored. Further research is needed to systematically evaluate their effectiveness and refine their implementation for music theory tasks.

To address these gaps, this study conducts a systematic comparative analysis of the performance of three LLMs (ChatGPT, Claude, Gemini) on music theory tasks across four music encoding formats (ABC, Humdrum, MEI, MusicXML). By evaluating multiple models and encoding representations, we aim to determine which approaches are most effective for in-context learning and chain-of-thought prompting in this domain.

To date and to the best of our knowledge, no systematic comparative studies of the suitability of different music encoding formats or the performance of different LLMs in music theory exist. This research seeks to address these gaps through the following research questions:
- What is the baseline performance of LLMs in music theory and how can in-context learning and chain-of-thought prompting enhance it?
- Which LLMs currently perform the best in these tasks?
- What music encoding formats are most conducive to these tasks?

By investigating the application of in-context learning and chain-of-thought prompting, this research seeks to address critical gaps in the pedagogical capabilities of LLMs in music theory. Through a comparative analysis of different models, music encoding formats, and prompting strategies, this study aims not only to evaluate the current state of LLMs in music theory, but also to explore their potential for students, educators, and future researchers.

## 2 BACKGROUND

This section provides an overview of the key theoretical and technical foundations relevant to this research. In Sections 2.1 and 2.2, we introduce in-context learning and chain-of-thought prompting, two advanced prompting techniques that are explored in this work. In Section 2.3, we discuss the strengths, limitations, and history of the four music encoding formats tested in this study (ABC, Humdrum, MEI, MusicXML). Since LLMs exhibit varying familiarity with different encoding formats, likely due to differences in their pretraining data, we investigate which formats are most suitable for in-context learning of music theory. Finally, in Section 2.4, we provide an overview of the Canadian Royal Conservatory of Music (RCM) examination system, which serves as our benchmark for evaluating LLM performance.

### 2.1 In-Context Learning

Introduced in 2020, in-context learning is a prompting technique in which an LLM learns patterns, rules, and concepts solely from the context provided within a prompt without fine-tuning the models or updating any underlying parameters. By presenting structured examples, explanations, and relevant background information, in-context learning allows models to recognize patterns and generalize solutions to new problems, particularly in reasoning-based tasks (Brown et al., 2020).

In-context learning currently takes advantage of large query context windows, which have recently grown to as many as 2 million tokens in the case of Google Gemini 1.5 Pro, a freely accessible model available online. To put this into perspective, 2 million tokens are equivalent to nearly five days of audio recordings or more than ten times the length of "War and Peace," a 1,440-page book (587,287 words). It is this ability to analyze massive amounts of data that makes in-context learning such a powerful tool (Gemini Team et al., 2024).

However, more is not always better. For example, the needle-in-a-haystack benchmark test for LLMs measures their ability to efficiently retrieve specific information from a vast amount of unrelated content (Machlab & Battle, 2024). As one would expect, performance declines as the 'haystack' grows—that is, as the prompt length increases. While newer models are rapidly improving on this benchmark, prompts designed for in-context learning are more like haystacks filled with needles, where much, if not all, of the content is relevant to the task. Unsurprisingly, when two prompts are of equal length, increasing the number of relevant details—or 'needles'—within the prompt leads to reduced performance, as the model needs to critically parse through more information to arrive at an answer.

For these reasons, an ideal prompt strikes a balance by providing enough background information to give the LLM the necessary context to learn the subject while avoiding excessive content that could overwhelm the model or dilute the relevance of key information.

While previous studies have explored machine learning optimization techniques for generating prompts (Hao et al., 2023), developing an effective query remains a highly iterative and experimental art (Bozkurt & Sharma, 2023). At first glance, the way machines learn may appear fundamentally different from human learning. However, because we interact with LLMs through a text-based interface, the process of teaching them more closely resembles communicating with a human learner than any other human-computer paradigm in the past.

## 2.2 Chain-of-Thought Prompting

Chain-of-thought prompting is a strategy that involves providing intermediate examples of the reasoning process needed to solve a problem, as illustrated in Figure 1. Unlike traditional machine learning frameworks, in which models are typically given many inputs and outputs without an explanation of the connection between them, chain-of-thought prompting is often used in conjunction with few-shot prompting, in which the model only has access to a small number of examples. A 3-shot chain-of-thought prompt, for instance, would include three step-by-step worked solutions to similar questions. Research has shown that this approach enhances performance by guiding models to use a more structured and systematic problem-solving methodology, leading to more rational and logical thinking (Wei et al., 2023; Zhang et al., 2022).

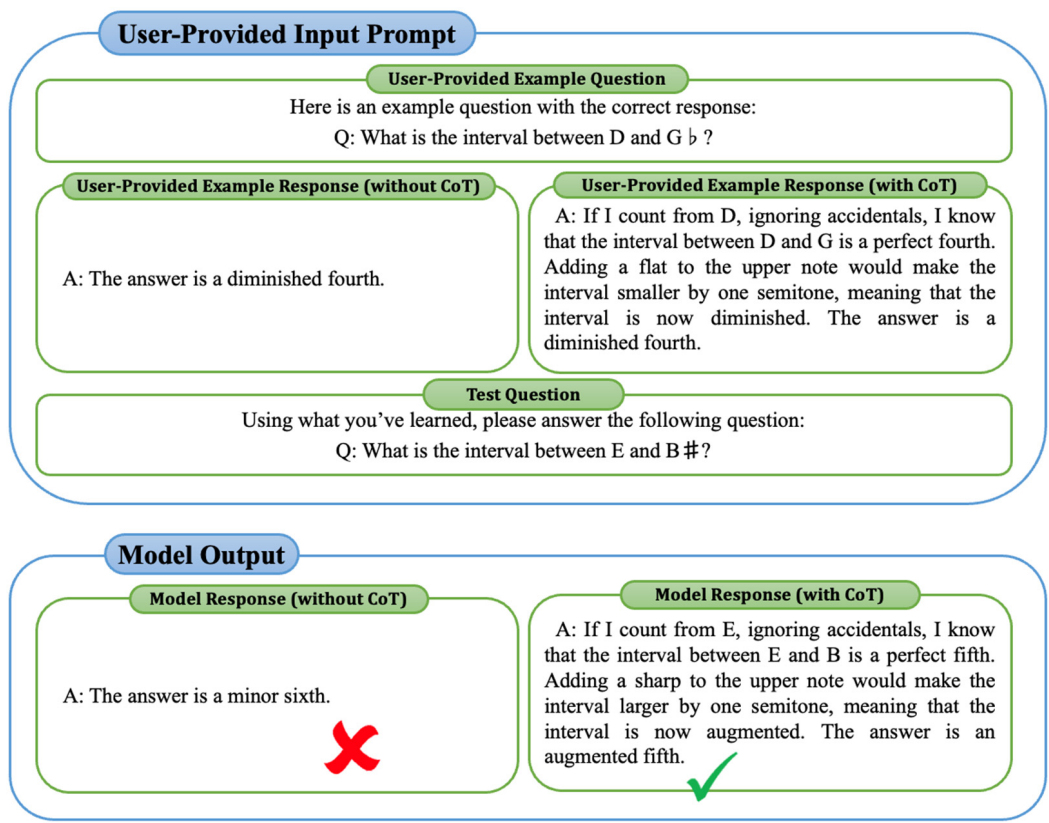

Figure 1: Chain-of-thought (CoT) prompting enables LLMs to solve more complex reasoning tasks. When the user provides chain-of-thought examples, as shown on the right side, the model learns the correct reasoning process, improving performance compared to without chain-of-thought, shown on the left.

## 2.3 Music Encoding Formats

While music encoding formats all ultimately share the same goal of representing musical scores in a text-based format, each has its own strengths and limitations. Some prefer efficiency while others aim to represent subtle aspects of scores in detail. In this study, four music encoding formats are investigated: ABC, Humdrum, MEI, and MusicXML.

Introduced in 1993, the ABC notation was originally developed for the concise representation of folk and traditional melodies, making it lightweight and easily readable (Walshaw, 2014). However, it is nevertheless powerful enough to encode most music scores, is actively maintained and developed, can easily be converted to other formats, and is human-readable, unlike many other more complex formats (Azevedo & Almeida, 2013).

Humdrum, on the other hand, is a text-based music toolkit developed primarily for computational musicology and analysis. A key strength of Humdrum is its extensibility—users can encode a wide variety of musical and non-musical data, including acoustic spectra and track index markers. A significant amount of non-Western music, from Aleutian to Zulu, has also been encoded using Humdrum. Distinctively, Humdrum arranges data vertically in columns called *spines*, allowing for different types of information, such as pitch, duration, harmonic analysis, or metadata, to be encoded in parallel while maintaining a logical temporal flow (Huron, 2002).

In 1999, at the University of Virginia, the Music Encoding Initiative (MEI) was developed to address the limitations of earlier music encoding systems, which were often project-specific and lacked generalizability. Like Humdrum, it has taken a community-driven approach and excels in representing a wide variety of musical notations. Its flexibility comes from its modular framework, which allows users to activate or deactivate encoding rules depending on the needs of each piece of music. Furthermore, MEI's structured XML-based design helps it integrate with other digital humanities tools, archives, and music analysis software (Roland et al., 2014).

Like MEI, MusicXML was developed as a common format for music notation. A major strength of MusicXML is its widespread industry adoption, with over 160 applications supporting it, including leading notation software like MuseScore and Sibelius. Notably, MusicXML uses semantic encoding, distinguishing between how music is notated, performed, and displayed, which makes it suitable for editing, playback, and archival purposes. However, unlike many other popular music encoding formats, MusicXML is optimized for Western musical notation, meaning that it lacks the flexibility that makes other systems ideal choices for early and non-Western music. Despite this, MusicXML remains an industry leader in digital sheet music interchange (Good, 2013).

Ultimately, all four have been used extensively in music information retrieval studies, making them ideal candidates for this research (E. Kijas et al., 2024; González Gutiérrez et al., 2022; Oliwa, 2008; Ratta & Daga, 2022; Ríos-Vila et al., 2024; Seeyo et al., 2024).

## 2.4 Royal Conservatory of Music

Founded in 1886 in Toronto, Canada, the Royal Conservatory of Music (RCM) is a Canadian music education institution with over five million alumni. In particular, the Certificate Program is an internationally recognized standardized system for music education and assessment for students of all ages and skill levels. It offers a sequential curriculum that integrates performance, technique, ear training, and sight reading. The program is structured with two Preparatory levels, followed by Levels 1–10, and the Associate (ARCT) and Licentiate (LRCM) Diplomas.

To accompany its practical curriculum, the RCM also offers examinations in music theory and history, which range from Level 5 to Analysis, Harmony & Counterpoint, History, and Keyboard Harmony, which correspond with the ARCT level.

In total, the Certificate Program is used by over 30,000 teachers and 500,000 students across North America [1] and provides a rigorous standardized testing framework for music education, making it an ideal benchmark to test musical understanding in LLMs.

---

[1] https://rcmusic.com/about-us/rcm-usa

Figure 2: Sample from Question 10 of the August 2024 RCM Level 6 Theory exam. Each LLM was tasked with analysing this multifaceted problem, drawing on knowledge from multiple areas of music theory.

## 3 METHODOLOGY

The musical understanding of three LLMs (ChatGPT, Claude, Gemini) with and without chain-of-thought prompting techniques were evaluated on a dataset of questions from the August 2024 Level 6 RCM Theory examination (see Figure 2). Musical excerpts were provided in four encoded formats: ABC, Humdrum, MEI, and MusicXML.

Models were queried through the API (Application Programming Interface) using Python (version 3.13.1) and were asked the questions both with and without context. The contextual prompts employed in-context learning and up to 3-shot chain-of-thought prompting and consisted of guides and examples for similar questions.

Due to their sheer size—up to 7,000 words—the complete prompts cannot be included in this document. However, they are publicly available and can be accessed through this project's GitHub repository.[2]

### 3.1 Large Language Models

The specific large language models tested were OpenAI's GPT-4o 2024-11-20, Anthropic's Claude 3.5 Sonnet 2024-10-22, and Google's Gemini 1.5 Pro. While Gemini 2.0 Flash was available at the time of the study and outperforms Gemini 1.5 Pro on many tasks, Gemini 1.5 Pro was chosen due to a December 2024 Google blog post, which indicated that Gemini 1.5 Pro significantly outperformed Gemini 2.0 Flash on long-context problems.[3]

### 3.2 Dataset

To construct the test dataset, questions from the Level 6 RCM Theory examination were used. Topics covered include pitch and notation, rhythm and meter, intervals, scales, chords and harmony, melody and composition, form and analysis, music terms and signs, and music history.

Proctored closed-book examinations took place across Canada and the United States at certified in-person examination centres at 9:30 am local time on 9–10 August 2024. The examination was out of 100 points and students had 120 minutes to answer 10 questions, most of which required writing music directly on the score. The national average for Level 6 exams written in 2024 was 88.0%.[4] A minimum grade of 60% is required to pass the examination.

---

[2] https://github.com/liampond/LLM-RCM

[3] https://blog.google/technology/google-deepmind/google-gemini-ai-update-december-2024/

### 3.3 Prompt Design

Prompts were iteratively and experimentally designed in collaboration with the LLMs themselves using questions from previous years. Preliminary investigation highlighted the variability of responses and the importance of subtle details in prompting. To improve comprehension and accuracy, pedagogical strategies traditionally used in human education were incorporated into the prompting design. These included asking the model to repeat its understanding of a topic before attempting a response and asking it to identify what additional information might be necessary to help it improve.

Conversely, machine-specific strategies were also tested. For instance, an interval lookup table was developed that would allow the model to effectively memorize the interval between every possible combination of two notes. However, this approach was abandoned as errors in accurately reading the table greatly limited its utility, indicating a more traditional prompting method would be more effective.

### 3.4 Prompt Structure

Models were first provided a system prompt (see Figure 3) to establish the model's role, behaviour, and response style. The query itself started by identifying the encoded file format and the premise of the prompt, followed by the encoded file itself, the relevant context—if applicable—and lastly the question.

The context was comprised of an outline of the question, followed by a guide or multiple guides of the relevant subject material, and then up to three chain-of-thought example problems on the same topic. For questions with multiple subcomponents involving integrating knowledge from different areas of music theory, preliminary investigation indicated performance degraded due to prompt length. In such cases, after the prompts had been optimized for brevity, the number of chain-of-thought example problems was reduced to two.

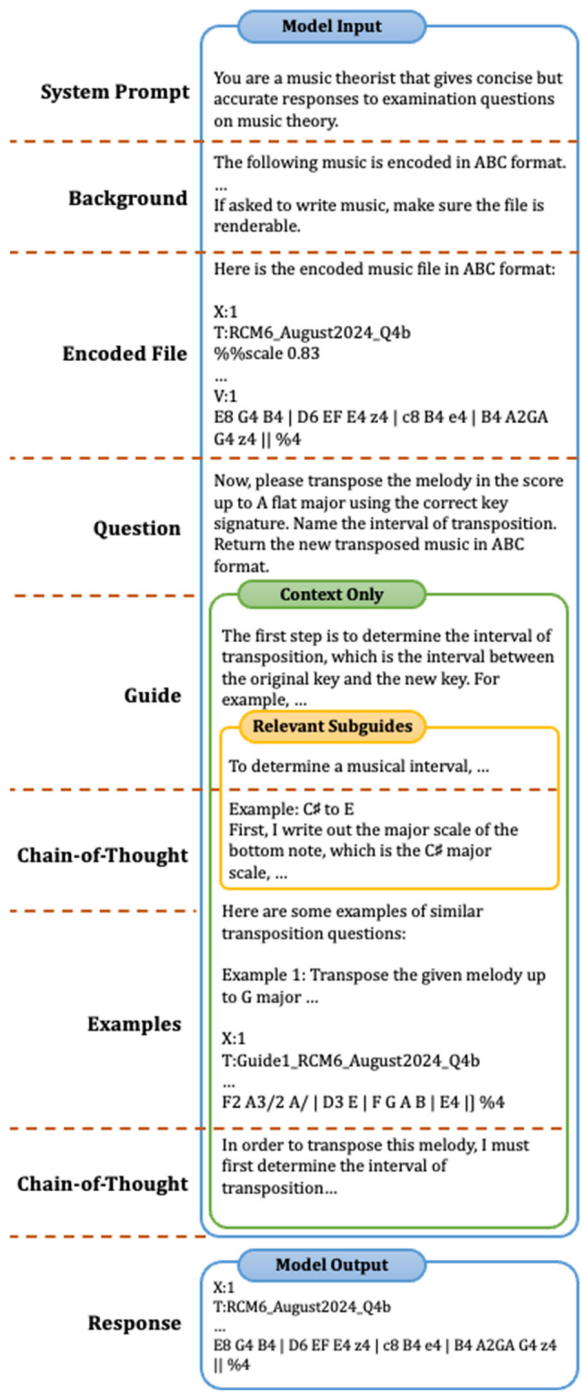

Figure 3: Structural overview of the prompt for Question 4b in ABC format. The model is queried twice, first without context, then with context (shown in green) using chain-of-thought prompting.

---

[4] https://www.rcmusic.com/learning/examinations/examination-averages

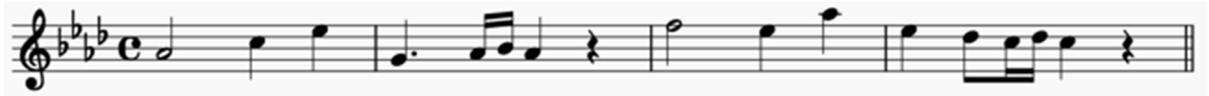

Figure 4: Music correctly produced in-context by Claude for Question 4b, which asks the candidate to transpose the given melody from C major to A♭ major. Encoded in the MEI format and rendered in MuseScore.

Table 1: Total evaluation results by datatype and LLM with and without context.

| Music Encoding Format | ABC | Humdrum | MEI | MusicXML |
|---|---|---|---|---|
| ChatGPT | 41% | 39% | 39% | **52%** |
| Claude | **42%** | **49%** | **44%** | 35% |
| Gemini | 39% | 19% | 30% | 34% |

(a) Total evaluation results for **no-context** prompts

| Music Encoding Format | ABC | Humdrum | MEI | MusicXML |
|---|---|---|---|---|
| ChatGPT | 48% (+7%) | 54% (+15%) | 60% (+21%) | 58% (+6%) |
| Claude | **57%** (+15%) | **74%** (+25%) | **75%** (+31%) | **74%** (+39%) |
| Gemini | 35% (-4%) | 40% (+21%) | 52% (+22%) | 54% (+20%) |

(b) Total evaluation results for **context** prompts with percent difference between context and no-context prompts

Models were asked to respond in prose or by generating music in the encoded format when applicable. While models were provided guides on music theory concepts, they were not explicitly taught the mechanics of each encoded file format, instead being given examples of properly formatted files through chain-of-thought prompting.

## 3.5 Evaluation

Each question in the examination was manually typeset in MuseScore (version 4.4.4), a free, open-source music notation program. MEI and MusicXML files were natively exported from MuseScore, while ABC and Humdrum files were converted from MusicXML using abcweb[5] and Humdrum's Music-XML-to-Humdrum Batch Converter,[6] respectively.

All questions were used except for Question 8, worth 10 points, which asked students to complete an unfinished melody based on functional chord symbols and compose an answer phrase to create a parallel period. Due to the subjectivity of evaluation, this question was omitted from the study.

All three LLMs (ChatGPT, Claude, Gemini) were evaluated on each of the remaining nine questions encoded in each of the four formats (ABC, Humdrum, MEI, MusicXML). Each model was first asked the examination question without any additional context to establish baseline performance and was then queried with the contextual prompts, which use in-context learning and chain-of-thought prompting to teach the model the subject material before asking the model the exam question.

The temperature parameter, which controls the randomness and creativity of responses, was set to zero for all models to encourage consistency. To minimize the effect of model variance, each prompt was queried three times, and the resulting scores were averaged. At the end, all responses were manually evaluated, and results were analysed in Microsoft Excel (version 16.93.1).

## 4 RESULTS AND DISCUSSION

Evaluation results, as shown in Table 1, indicate moderate variation in performance across LLMs and music encoding formats. However, contextual prompts consistently improved accuracy across all models and formats, with the exception of Gemini on the ABC format, whose performance decreased from 39% to 35%, likely due to variability among outputs.

Results were similar across music encoding formats, although performance was generally worse on ABC than with Humdrum, MEI, or MusicXML. In terms of models, Claude significantly outperformed both ChatGPT and Gemini with context and nearly matched ChatGPT's performance without context.

---

[5] https://wim.vree.org/js/xml2abc-js.html

[6] https://github.com/humdrum-tools/musicxml-batch-converter

If the models were to have written the examination as a human student would, none of the models would have met the passing threshold of 60% without the help of the contextual prompts. However, with context, Claude would have received Honors (70–79%) on three of the four encoding formats. Notably, neither of the other two models would have passed, with the exception of ChatGPT, which would meet the passing grade with MEI.

As illustrated by the detailed breakdown in Table A1 and Table A2 in the Appendix, the effectiveness of contextual prompting varied significantly depending on the type of question. Substantial improvements were observed for problems related to intervals, scales, transposition, and cadences, where the structured guides and chain-of-thought examples were able to effectively teach the machine these concepts.

However, questions related to chords showed limited improvement, suggesting that the LLMs struggled with the multi-step sequential nature of these problems. Rhythmic understanding proved even more challenging, as attempts to teach rests and rhythmic groupings were unsuccessful. Similarly, this was likely due to the inherent complexity of the topic, which requires an understanding of time signatures, beat patterns, and a hierarchy of rules regarding the situations in which various rests may or may not be combined. Furthermore, this issue could be exacerbated by the abundance of poorly notated freely available music on the web, which the LLMs were likely trained on.

Unsurprisingly, even without context, all LLMs achieved perfect scores on the knowledge-based questions which asked students to match opposite musical terms, like *senza* and *con*, and provide terminology based on a short description, such as identifying the *recapitulation* as the third section in sonata form.

Finally, of the incorrect responses, many were due to models returning corrupted or unreadable files, suggesting that musical comprehension may have been sufficient for the tasks but was limited by the model's understanding of the encoded formats. While these files were usually consistently unreadable across the three trials, files were sometimes corrupted for only some of the trials and readable or partially readable when queried again.

Ultimately, while LLMs in their current states would need improvements before being reliably adopted in educational settings, this work provides a foundation for more in-depth research. LLMs proficient in music theory have the potential to revolutionize both teaching and learning practices. In the future, these models could serve as on-demand digital tutors, offering instantaneous explanations of complex concepts while assisting students in self-inquiry by providing personalized feedback tailored to their individual needs.

In addition, these systems could help educators by generating focused practice questions, problem sets, and mock examinations that align with standardized curricula or learning objectives. This would not only alleviate the workload of increasingly overburdened teachers but also allow them to spend a greater percentage of their time interacting with students.

## 5 CONCLUSION AND FUTURE WORK

This study explored the capabilities of Large Language Models (LLMs) in learning music theory through in-context learning and chain-of-thought prompting. Using Royal Conservatory of Music examination questions as a benchmark, we systematically evaluated ChatGPT, Claude, and Gemini across four symbolic music encoding formats (ABC, Humdrum, MEI, and MusicXML).

Results demonstrated that contextual prompting significantly enhanced performance, particularly for questions related to intervals, scales, transposition, and cadences. However, chord analysis showed minimal improvement, and we were unable to teach the models rests and rhythmic groupings. Despite these limitations, the findings indicate that LLMs can be effectively guided to improve music theory reasoning through carefully structured prompts that teach the models step-by-step, drawing notable parallels between human and machine learning processes.

Future work should explore more advanced music theory concepts, of which there is an abundance. Because LLMs are rapidly evolving, these results could quickly become out of date and should be reinvestigated as models improve and in-context learning becomes more efficient. In addition, the performance of other LLMs like DeepSeek, LLaMA, and Mistral should be explored.

## ACKNOWLEDGEMENTS

This project is supported in part by the Social Sciences and Humanities Research Council of Canada (SSHRC 895-2022-1004) and the Fonds de Recherche du Québec (FRQSC SE3-303927).

# APPENDIX

Table 2: Detailed breakdown of results **without context** by question. Each LLM was evaluated twelve times per question (three trials per question across four encoding formats).

| Music Encoding Format | ABC | Humdrum | MEI | MusicXML |
|---|---|---|---|---|
| 1. Intervals | 26.7% | 46.7% | 43.3% | 60.0% |
| 2. Rests and Rhythms | 0.0% | 10.0% | 0.0% | 0.0% |
| 3. Scales | 33.3% | 26.7% | 0.0% | 61.7% |
| 4. Transposition | 25.0% | 0.0% | 0.0% | 0.0% |
| 5. Chords | 0.0% | 0.0% | 0.0% | 45.0% |
| 6. Cadences | 60.0% | 80.0% | 73.3% | 83.3% |
| 7. Keys, Rhythms, Chords | 53.3% | 56.7% | 70.0% | 56.7% |
| 9. Terminology and History | 100.0% | 100.0% | 100.0% | 100.0% |
| 10. Integrated Knowledge | 66.7% | 33.3% | 68.3% | 63.3% |

(a) Detailed breakdown of results **without context** for ChatGPT

| Music Encoding Format | ABC | Humdrum | MEI | MusicXML |
|---|---|---|---|---|
| 1. Intervals | 30.0% | 63.3% | 60.0% | 26.7% |
| 2. Rests and Rhythms | 0.0% | 0.0% | 0.0% | 0.0% |
| 3. Scales | 0.0% | 0.0% | 33.3% | 0.0% |
| 4. Transposition | 55.0% | 10.0% | 0.0% | 0.0% |
| 5. Chords | 0.0% | 0.0% | 0.0% | 0.0% |
| 6. Cadences | 90.0% | 93.3% | 90.0% | 66.7% |
| 7. Keys, Rhythms, Chords | 40.0% | 80.0% | 36.7% | 60.0% |
| 9. Terminology and History | 100.0% | 100.0% | 100.0% | 100.0% |
| 10. Integrated Knowledge | 60.0% | 90.0% | 80.0% | 63.3% |

(b) Detailed breakdown of results **without context** for Claude

| Music Encoding Format | ABC | Humdrum | MEI | MusicXML |
|---|---|---|---|---|
| 1. Intervals | 10.0% | 0.0% | 20.0% | 40.0% |
| 2. Rests and Rhythms | 0.0% | 0.0% | 0.0% | 6.7% |
| 3. Scales | 40.0% | 0.0% | 0.0% | 0.0% |
| 4. Transposition | 26.7% | 10.0% | 0.0% | 0.0% |
| 5. Chords | 0.0% | 0.0% | 0.0% | 0.0% |
| 6. Cadences | 80.0% | 20.0% | 60.0% | 70.0% |
| 7. Keys, Rhythms, Chords | 40.0% | 0.0% | 30.0% | 30.0% |
| 9. Terminology and History | 100.0% | 100.0% | 100.0% | 100.0% |
| 10. Integrated Knowledge | 53.3% | 36.7% | 60.0% | 60.0% |

(c) Detailed breakdown of results **without context** for Gemini

Table 3: Detailed breakdown of **in-context** results by question. Each LLM was evaluated twelve times per question (three trials per question across four encoding formats).

| Music Encoding Format | ABC | Humdrum | MEI | MusicXML |
|---|---|---|---|---|
| 1. Intervals | 36.7% | 40.0% | 50.0% | 60.0% |
| 2. Rests and Rhythms | 0.0% | 10.0% | 10.0% | 10.0% |
| 3. Scales | 68.3% | 78.3% | 71.7% | 75.0% |
| 4. Transposition | 40.0% | 20.0% | 81.7% | 6.7% |
| 5. Chords | 16.7% | 30.0% | 35.0% | 40.0% |
| 6. Cadences | 61.7% | 80.0% | 31.7% | 71.7% |
| 7. Keys, Rhythms, Chords | 60.0% | 73.3% | 83.3% | 86.7% |
| 9. Terminology and History | 100.0% | 100.0% | 100.0% | 100.0% |
| 10. Integrated Knowledge | 50.0% | 50.0% | 76.7% | 70.0% |

(a) Detailed breakdown of **in-context** results for ChatGPT

| Music Encoding Format | ABC | Humdrum | MEI | MusicXML |
|---|---|---|---|---|
| 1. Intervals | 53.3% | 73.3% | 90.0% | 100.0% |
| 2. Rests and Rhythms | 10.0% | 10.0% | 10.0% | 0.0% |
| 3. Scales | 85.0% | 90.0% | 100.0% | 100.0% |
| 4. Transposition | 100.0% | 100.0% | 75.0% | 50.0% |
| 5. Chords | 30.0% | 45.0% | 70.0% | 70.0% |
| 6. Cadences | 26.7% | 100.0% | 100.0% | 100.0% |
| 7. Keys, Rhythms, Chords | 50.0% | 66.7% | 53.3% | 66.7% |
| 9. Terminology and History | 100.0% | 100.0% | 100.0% | 100.0% |
| 10. Integrated Knowledge | 56.7% | 76.7% | 73.3% | 80.0% |

(b) Detailed breakdown of **in-context** results for Claude

| Music Encoding Format | ABC | Humdrum | MEI | MusicXML |
|---|---|---|---|---|
| 1. Intervals | 13.3% | 26.7% | 53.3% | 30.0% |
| 2. Rests and Rhythms | 0.0% | 10.0% | 0.0% | 0.0% |
| 3. Scales | 20.0% | 85.0% | 86.7% | 95.0% |
| 4. Transposition | 0.0% | 0.0% | 20.0% | 0.0% |
| 5. Chords | 30.0% | 30.0% | 25.0% | 40.0% |
| 6. Cadences | 66.7% | 31.7% | 70.0% | 71.7% |
| 7. Keys, Rhythms, Chords | 33.3% | 46.7% | 66.7% | 83.3% |
| 9. Terminology and History | 100.0% | 100.0% | 100.0% | 100.0% |
| 10. Integrated Knowledge | 53.3% | 30.0% | 50.0% | 66.7% |

(c) Detailed breakdown of **in-context** results for Gemini